\documentclass{webofc}
\usepackage[varg]{txfonts}   
\usepackage{xcolor}
\begin{document}
\title{Tracing baryon and electric charge transport\\ in isobar collisions}
%

\author{\firstname{Gregoire} \lastname{Pihan}\inst{1}\fnsep\thanks{\email{GregoirePihan@wayne.edu} (Speaker)} \and
        \firstname{Akihiko} \lastname{Monnai}\inst{2} \and
        \firstname{Bj\"orn} \lastname{Schenke}\inst{3}\and
        \firstname{Chun} \lastname{Shen}\inst{1,4}
}

\institute{Department of Physics and Astronomy, Wayne State University, Detroit, Michigan 48201, USA
\and
           Department of General Education, Faculty of Engineering, Osaka Institute of Technology, Osaka 535-8585, Japan 
\and
           Physics Department, Brookhaven National Laboratory, Upton, NY 11973, USA
\and
           RIKEN BNL Research Center, Brookhaven National Laboratory, Upton, NY 11973, USA
}

\abstract{%
  It is of fundamental interest to understand the carrier of conserved quantum charges within protons and nuclei at high energy. Preliminary data from isobar collisions at RHIC reveal a scaled net-baryon to net-electric charge ratio ($B/\Delta Q \times \Delta Z/A$) at midrapidity between 1.2 and 2, consistent with string junction model predictions.
  Here, we compute the initial stage scaled net-baryon to net-electric charge ratio for isobar collisions. Our model incorporates a realization of the string junction model and models the nuclear structure. Our predictions identify the baseline expectations for such measurement and quantify the impact of the nuclear structure. 
}
\maketitle
\section{Introduction}
\label{intro}
A gauge-invariant approach to constructing a proton's baryon charge is to connect its three valence quarks through a three-gluon vertex known as the string junction \cite{Montanet:1980te}. It was later proposed that the effective carrier of baryon number within a baryon could be the string junction itself \cite{Kharzeev:1996sq}. In that case, it is anticipated that in relativistic nuclear collisions the net baryon number ($B$) will experience more stopping at initial impacts than the net electric charge number ($Q$), assigned to the valence quarks. The precise measurement of the baryon-to-electric-charge stopping ratio in isobar collisions (Ru+Ru: $A,Z=96,44$, Zr+Zr: $A,Z=96,40$) at RHIC \cite{STAR:2021mii} offers an unprecedented opportunity to test this assumption experimentally. Any deviation from global charge conservation ($Q\sim Z/A\times B$) at mid-rapidity may signify distinct longitudinal dynamics for $B$ and $Q$ during the collision or the possibility that $B$ is not carried by the valence quarks, supporting the string junction model~\cite{Lewis:2022arg}.
In this study, we compute the initial conditions for  $\sqrt{s_{\text{NN}}} = 200$ GeV isobar collisions, utilizing a \textsc{3D-Glauber} model \cite{Shen:2017bsr, Shen:2022oyg}. These initial conditions can then be used in the \textsc{iEBE-MUSIC} framework to produce final-state particle observables to be compared to the STAR measurements. Understanding the initial conditions for isobar collisions is an essential step in studying the baryon-to-electric charge-stopping ratio. Indeed, all key features observed at the initial stage are expected to be seen at the final stage if the longitudinal dynamics are similar for different flavors of conserved charges.

\section{The Theoretical Framework}

In the \textsc{3D-Glauber} model, nucleon-nucleon collisions decelerate incoming partons and produce strings as energy-momentum source currents for the hydrodynamic medium~\cite{Shen:2017ruz, Shen:2017fnn, Shen:2017bsr}. The baryon numbers from the incoming nucleons can either remain in the wounded beam remnants or fluctuate inside the string pulled by the colliding parton pairs~\cite{Shen:2022oyg}. If the baryon number is assigned to the string, its rapidity $y^{B}_{P/T}$ follows a probability distribution given by
\begin{equation}
    P(y^{X}_{P/T}) = (1 - \lambda_X) y_{P/T} + \lambda_X \frac{e^{(y^{X}_{P/T} - (y_P + y_T)/2)/2}}{4 \sinh{((y_P - y_T)/4)}},
    \label{eq:GJprobability}
\end{equation}
where $P$ and $T$ represent the projectile and target, respectively and $X=B$~\cite{Kharzeev:1996sq, Shen:2022oyg}.
Following this probability, the baryon number may be placed at the string end with a probability of ${(1-\lambda_B)}$, consistent with the assumption of baryon number carried by valence quarks, or toward mid-rapidity based on predictions from the string junction, with a probability $\lambda_B$. The STAR measurements on midrapidity net-proton yields in Au+Au collisions in the Beam Energy Scan program favor $\lambda_B = 0.2$~\cite{STAR:2017sal, Shen:2022oyg}.

This work considers the electric charge number to fluctuate toward mid-rapidity following the same probability as in Eq.~\eqref{eq:GJprobability} with $X=Q$ and an independent model parameter $\lambda_Q$. While this is not motivated by a specific physical picture for electric charge stopping, it enables control over the baryon-to-electric-charge stopping ratio at the initial stage. If $\lambda_Q = 0$, the electric charge is associated with valence quarks. Suppose $\lambda_Q = \lambda_B$, both the baryon and the electric charge numbers stop in the same way. Throughout the remainder of this work, we will consider the baseline calculations for the $\lambda_Q = \lambda_B$ case.
This procedure yields the initial rapidity distribution of baryon and electric charge numbers for isobar collisions.

The \textsc{3D-Glauber} model incorporates modeling nuclear structure through deformed Woods-Saxon distributions for protons and neutrons, including possible finite neutron skins for the colliding nuclei. The Woods-Saxon densities in a nucleus are expressed in spherical coordinates $(r, \theta, \phi)$ as
\begin{align}
\rho_i(r,\theta, \phi) = \frac{\rho_{0,i}}{1+e^{[r-R_i(\theta, \phi)]/a_i}},
\end{align}
where $i=p, n$ denotes protons and neutrons, $\rho_{0,i}$ is the central density and $a_i$ is the diffusivity parameter. Special deformation can be introduced via the radius parameter $R_i$ with spherical harmonic functions $Y_{lm}(\theta, \phi)$ as
\begin{align}
R_i(\theta, \phi) = R_{0,i}\Big[ 1 + \beta^{WS}_2 [\cos(\gamma) Y_{20}(\theta, \phi) + \sin(\gamma) Y_{22}(\theta, \phi)] + \beta^{WS}_3 Y_{30}(\theta, \phi) + \beta^{WS}_4 Y_{40}(\theta, \phi) \Big],
\end{align}
where $\beta^{WS}_{2,3,4}$ and $\gamma$ are the deformation parameters that define the nucleus shape. The Ruthenium (Ru) and Zirconium (Zr) nuclei exhibit different shapes, and the parameters used in this study are summarized in Table~\ref{tab-1} based on a recent Density Functional Theory calculation~\cite{Xu:2021vpn}. The neutron skin is modeled with $R_{0,n} = R_{0, p} + dR$ and $a_n = a_p + da$.

\begin{table}[b!]
\vspace{-0.1cm}
\centering
\caption{Woods-Saxon density parameters for the Ruthenium and Zirconium nuclei~\cite{Xu:2021vpn}.}
\label{tab-1}
\begin{tabular}{|c|c|c|c|c|c|c|c|c|}
\hline
& $R_{0,p}$ (fm) & $a_p$  (fm) & $\gamma$ & $\beta^{WS}_2$ & $\beta^{WS}_3$ & $\beta^{WS}_4$ & $da$  (fm) & $dR$  (fm) \\ \hline
Ruthenium & 5.09 & 0.46 & 0 & 0.16 & 0 & 0 & 0.01 & 0.015 \\ \hline
Zirconium & 5.02 & 0.52 & 0 & 0.06 & 0.2 & 0 & 0.05 & 0.1 \\ \hline
\end{tabular}
\end{table}

\section{Results}

\begin{figure}[t!]
\centering
\includegraphics[width=0.48\linewidth]{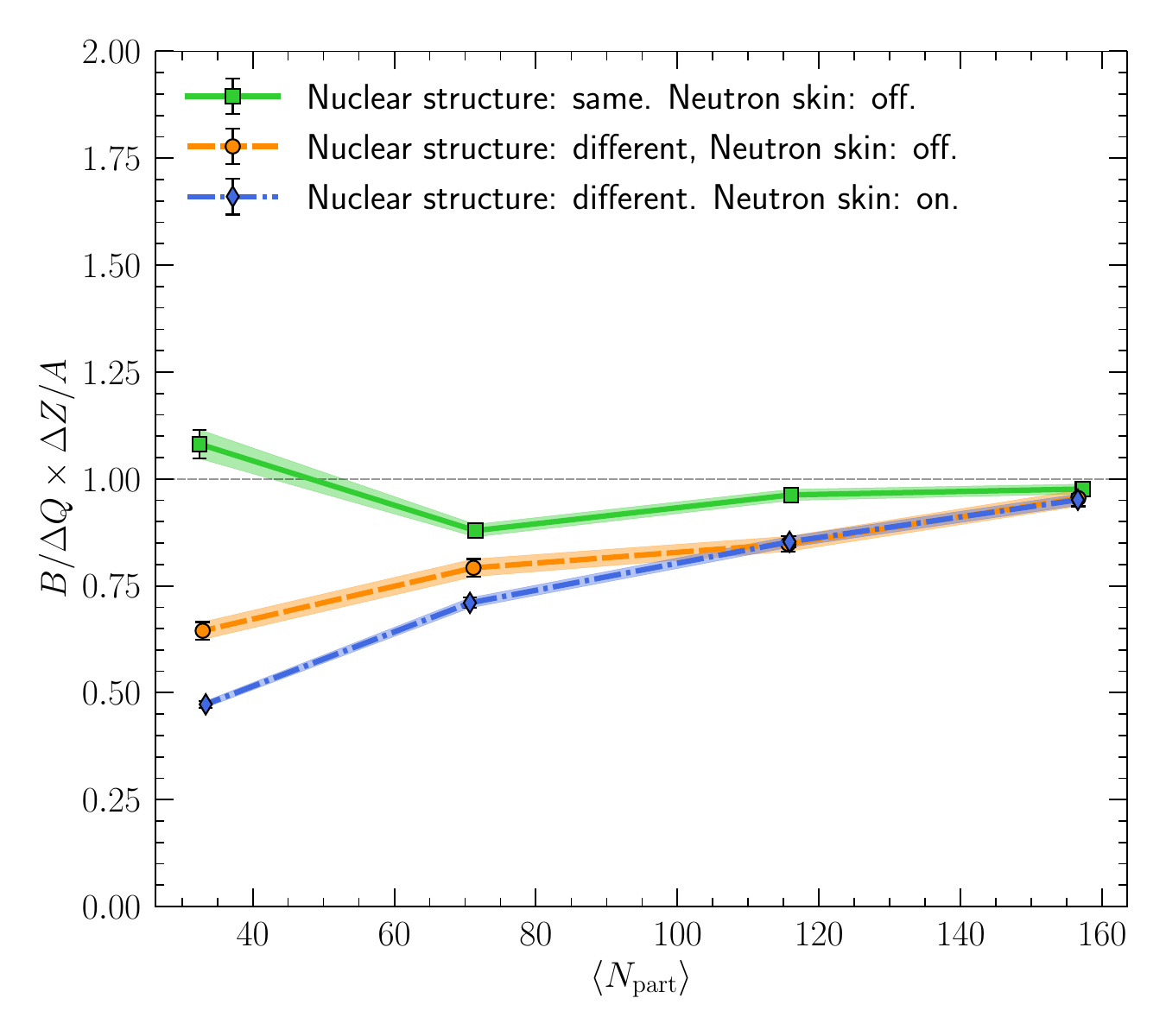}
\caption{The baryon to electric charge stopping ratio from the \textsc{3D-Glauber} model with $\lambda_B = \lambda_Q = 0.2$ for different nuclear structure configurations.}
\label{fig-1}       
\end{figure}

The \textsc{3D-Glauber} model provides the initial-state baryon and electric charge rapidity distributions. We then derive the stopping ratio observable as described in Ref.~\cite{Lewis:2022arg}. The isobar stopping ratio, denoted as $B/\Delta Q \times \Delta Z/A$, involves the net baryon number $B$ measured at mid-rapidity ($-0.5 < y < 0.5$) in Ru+Ru collisions, and $\Delta Q = Q_{\text{Ru}} - Q_{\text{Zr}}$, representing the net electric charge difference between Ru+Ru and Zr+Zr collisions at mid-rapidity. According to global charge conservation, if the baryon and the electric charge number stop equivalently, this ratio is expected to be close to unity.

We consider the following three scenarios to quantify the effects of the nuclear structure. First, the nuclear structure and neutron skins of Ru and Zr nuclei are defined by the parameters in Table~\ref{tab-1}. Second, the nuclear deformation is considered for both nuclei, but neutron skins are removed (${ \rm d}a = {\rm d}R = 0$ fm). Third, the nuclear structure is set to be identical for both nuclei (the Ru parameters in Table~\ref{tab-1}) and no neutron skin. Figure~\ref{fig-1} presents the predicted baryon-to-electric-charge ratio from the \textsc{3D-Glauber} model for the equal stopping configuration ($\lambda_B = \lambda_Q = 0.2$) for the three nuclear structure configurations described above.

In the first scenario (depicted by the blue dashed-dotted line), the ratio consistently remains below unity, and it increases with $N_{\text{part}}$, demonstrating that even in an equal stopping situation, the baseline for the ratio is not expected to be equal to one and flat in centrality. The reasons for the observed behavior become clearer when examining the other two cases. When the Ru and Zr nuclei are set to have identical shapes (illustrated by the green solid line), the ratio exhibits a flatter trend and is close to unity. This observation underscores the nuclear structure's significant impact on the ratio's centrality dependence in the full calculation. When we remove the neutron skin while keeping the different nuclear deformation in Ru and Zr nuclei, the result is close to the full calculation in central collisions. The different sizes of neutron skins in Ru and Zr nuclei give a 20\% smaller ratio in the peripheral collisions, at which collisions are more dominant by the proton-to-neutron ratio at the edge of the colliding nuclei.

In an equal stopping configuration, where both baryon and electric charge have an equal probability of stopping at the initial stage, the isobar-stopping ratio significantly deviates from unity because of the difference in nuclear structure between the Ru and Zr nuclei. Suppose an experimental study of the isobar-stopping ratio reveals an increasing centrality dependence. A comparison with our predictions, considering the full nuclear shape, is essential to disentangle the effects from flavor-dependent stopping powers at the initial stage. 

\section{Conclusions}

In this study, we used the \textsc{3D-Glauber} initial-state model to simulate isobar collisions at $\sqrt{s_{\text{NN}}} = 200$ GeV at RHIC, with a specific emphasis on tracing the flavor-dependence dynamics for conserved charges.
Our findings reveal that, even with an equal probability of stopping for baryon and electric charge, the expected ratio deviates from unity during the initial stage, primarily attributed to nuclear structure effects. 

\section*{Acknowledgments}
This work is supported by the U.S. Department of Energy (DOE), Office of Science, Office of Nuclear Physics, under DOE Contract No. DE-SC0012704 (B.P.S.) and Award No. DE-SC0021969 (G.P. and C.S.). A.M. was partially supported by JSPS KAKENHI Grant Number JP19K14722.
C.S. acknowledges a DOE Office of Science Early Career Award.
This research was done using computational resources provided by the Open Science Grid (OSG)~\cite{Pordes:2007zzb, Sfiligoi:2009cct}, which is supported by the NSF award \#2030508 and \#1836650.

\bibliography{references}

\begin{thebibliography}{12}

\bibitem{Montanet:1980te}
L.~Montanet, G.C. Rossi, G.~Veneziano, Phys. Rept. \textbf{63}, 149 (1980)

\bibitem{Kharzeev:1996sq}
D.~Kharzeev, Phys. Lett. B \textbf{378}, 238 (1996), \texttt{nucl-th/9602027}

\bibitem{STAR:2021mii}
M.~Abdallah et~al. (STAR), Phys. Rev. C \textbf{105}, 014901 (2022),
  \texttt{2109.00131}

\bibitem{Lewis:2022arg}
N.~Lewis, W.~Lv, M.A. Ross, C.Y. Tsang, J.D. Brandenburg, Z.W. Lin, R.~Ma,
  Z.~Tang, P.~Tribedy, Z.~Xu (2022), \texttt{2205.05685}

\bibitem{Shen:2017bsr}
C.~Shen, B.~Schenke, Phys. Rev. C \textbf{97}, 024907 (2018),
  \texttt{1710.00881}

\bibitem{Shen:2022oyg}
C.~Shen, B.~Schenke, Phys. Rev. C \textbf{105}, 064905 (2022),
  \texttt{2203.04685}

\bibitem{Shen:2017ruz}
C.~Shen, G.~Denicol, C.~Gale, S.~Jeon, A.~Monnai, B.~Schenke, Nucl. Phys. A
  \textbf{967}, 796 (2017), \texttt{1704.04109}

\bibitem{Shen:2017fnn}
C.~Shen, B.~Schenke, PoS \textbf{CPOD2017}, 006 (2018), \texttt{1711.10544}

\bibitem{STAR:2017sal}
L.~Adamczyk et~al. (STAR), Phys. Rev. C \textbf{96}, 044904 (2017),
  \texttt{1701.07065}

\bibitem{Xu:2021vpn}
H.j. Xu, H.~Li, X.~Wang, C.~Shen, F.~Wang, Phys. Lett. B \textbf{819}, 136453
  (2021), \texttt{2103.05595}

\bibitem{Pordes:2007zzb}
R.~Pordes et~al., J. Phys. Conf. Ser. \textbf{78}, 012057 (2007)

\bibitem{Sfiligoi:2009cct}
I.~Sfiligoi, D.C. Bradley, B.~Holzman, P.~Mhashilkar, S.~Padhi, F.~Wurthwrin,
  WRI World Congress \textbf{2}, 428 (2009)

\end{thebibliography}

\end{document}